\documentclass[aps,twocolumn,prd,dvipsnames,usenames,floatfix,superscriptaddress,nofootinbib]{revtex4-1}
\pdfoutput=1

\usepackage{footnote}
\makesavenoteenv{tabular}
\makesavenoteenv{table}
\usepackage[utf8x]{inputenc}
\usepackage{units}
\usepackage[english]{babel}
\usepackage{xspace}
\usepackage{paralist}
\usepackage{amsmath} 
\usepackage{amssymb} 
\usepackage{graphicx}
\usepackage{bm}
\usepackage{tabularx}
\usepackage[final,textsize=footnotesize,color=ForestGreen!05]{todonotes}
\PassOptionsToPackage{unicode}{hyperref}
\PassOptionsToPackage{bookmarks=false}{hyperref}
\usepackage{hyperref}

\newcommand\myshade{85}
\colorlet{mylinkcolor}{violet}
\colorlet{mycitecolor}{Aquamarine}
\colorlet{myurlcolor}{YellowOrange}

\hypersetup{
 linkcolor  = mylinkcolor,
 citecolor  = mycitecolor!\myshade!black,
 urlcolor   = myurlcolor!\myshade!black,
colorlinks = true,
}

\newcommand{\myparagraph}[1]{\bigskip\noindent\textbf{#1}}
\newcommand{\Br}{\mathrm{Br}}
\newcommand{\decay}{\mathrm{dec}}
\newcommand{\events}{\mathrm{events}}
\renewcommand{\min}{\mathrm{min}}
\renewcommand{\max}{\mathrm{max}}
\begin{document}

\title{Probing new physics with displaced vertices: muon tracker at CMS}

 \author{Kyrylo~Bondarenko}
 \author{Alexey~Boyarsky}
 \author{Maksym~Ovchynnikov}
 \affiliation{Intituut-Lorentz, Leiden University, Niels Bohrweg 2, 2333 CA Leiden, The Netherlands} 
 \author{Oleg~Ruchayskiy}
 \affiliation{Discovery Center, Niels Bohr Institute, Copenhagen University, Blegdamsvej 17, DK-2100 Copenhagen, Denmark} 
 \author{Lesya~Shchutska}
 \affiliation{Institute of physics, Laboratoire de physique des hautes énergies (LPHE), École polytechnique fédérale de Lausanne (EPFL), BSP 625, Rte de la Sorge, CH-1015 Lausanne, Switzerland}

\begin{abstract}
Long-lived particles can manifest themselves at the LHC via ``displaced vertices'' -- several charged tracks originating from a position separated from the proton interaction point by a macroscopic distance. Here we demonstrate the potential of the muon trackers at the CMS experiment for displaced vertex searches. We use heavy neutral leptons and Chern-Simons portal as two examples of long-lived particles for which the CMS muon tracker can provide essential information about their properties.
\end{abstract}

\maketitle 

\section{Introduction}
\label{sec:introduction}
The Standard model (SM) of particle physics is very successful in describing known elementary particles and their interactions. However it does not incorporate a number of observations in particle physics and cosmology: neutrino oscillations, generation of baryon asymmetry of the Universe, the nature of dark matter. 
These phenomena imply that new particles \textit{beyond the Standard Model}  exist.
Many extensions of the SM introduce new particles with macroscopic decay lengths ($c\tau > 1$~cm)~\cite{Giudice:1998bp,Barbier:2004ez,Asaka:2005an,Asaka:2005pn,Chacko:2005pe,Strassler:2006im,Strassler:2006ri,Burdman:2006tz,Cai:2008au,Baumgart:2009tn,Kaplan:2009ag,Chan:2011aa,Fan:2011yu,Arvanitaki:2012ps,Fan:2012jf,ArkaniHamed:2012gw,Dienes:2012yz,Csaki:2013jza,Chacko:2015fbc,Batell:2016zod,Fradette:2017sdd,Dienes:2011ja}. Such particles are often called \textit{long-lived particles} (LLPs).

\begin{figure}[h!]
    \centering
    \includegraphics[width=0.45\textwidth]{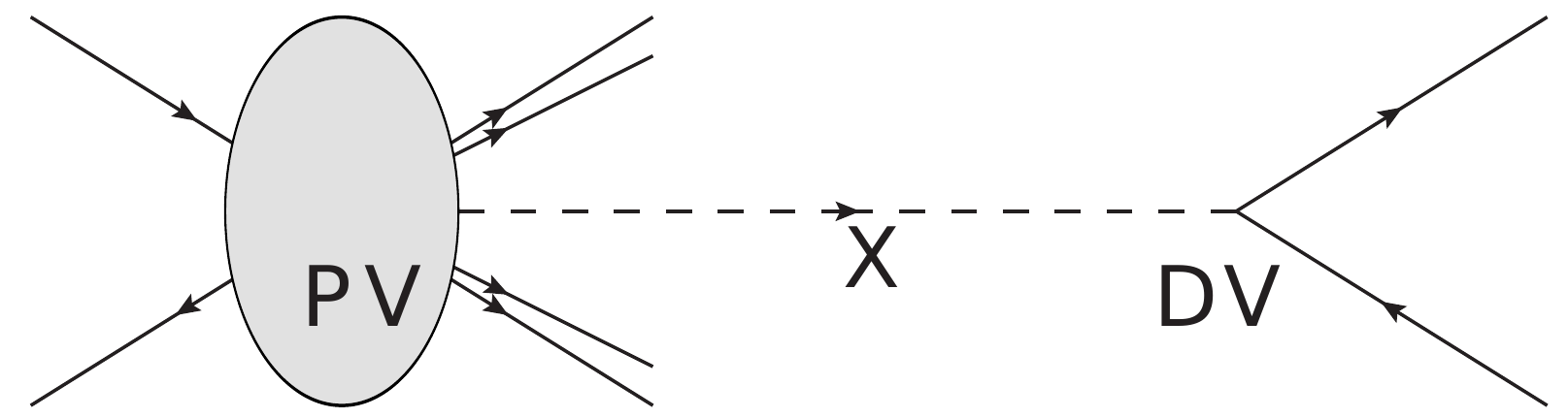}
    \caption{Schematic diagram of an LLP search at particle physics experiments. The LLP $X$ produced at the production vertex (PV) travels a macroscopic distance and then gives rise to a displaced decay vertex (DV).}
    \label{fig:displaced-vertex}
\end{figure}

LLPs can be copiously produced at colliders. Their decays present a distinct experimental signature --  the position of an LLP decay vertex is highly displaced from its production vertex (PV), see Fig.~\ref{fig:displaced-vertex}. 
Therefore, in order to probe the parameter space of the LLP, experiments
must have large length of the decay volume.

Searches for the  displaced vertex (DV) events  have been performed at the LHC in the past, including searches for displaced hadronic jets~\cite{Aaboud:2018aqj,Aaboud:2017iio,Sirunyan:2018pwn,Sirunyan:2017jdo,Sirunyan:2018vlw}, dimuon vertices~\cite{Aaboud:2018jbr,CMS:2015pca} and decays into specific final states (for example, the process $B^{\pm} \to \mu^{\pm}\mu^{\pm}\pi^{\mp}$ with one prompt and one displaced muon at LHCb~\cite{Aaij:2014aba}). 

In this paper, we propose a scheme of DV searches exploiting the muon trackers of the CMS experiment. Its advantage is large length of the decay volume, which is essential to probe the parameter space of the LLPs with the decay lengths about $ 1\text{ meter}$ or larger.

The use of the muon chambers to reconstruct di-muon DV signatures  has been explored in the past in~\cite{Bobrovskyi:2011vx,Bobrovskyi:2012dc,Izaguirre:2015pga} and recently in~\cite{Boiarska:2019jcw}. Ref.\ \cite{Drewes:2019fou} that appeared when this work was at its final stage employed the event selection criteria that may be too optimistic with regard to the background estimates. 
Ref.~\cite{CMS:2015pca}  explored a potential of the CMS muon chambers \textit{alone} to reconstruct dimuon DV. 
This search however, was constructed to be much more general, and hence could not profit from the presence of a prompt lepton in the event. This necessarily implied much more stringent cuts on $p_T$ of either of the two muons in the muon tracker since these muons were used to record an event by a trigger, and therefore lower sensitivity. 

This paper is organized as follows. In Sec.~\ref{sec:cms-muon-tracker} we discuss the ``long DV search scheme'', selection criteria and conditions for an LLP model under which the scheme is the most efficient. In Sec.~\ref{sec:examples} we discuss the potential of the scheme for specific models --- the neutrino portal and the Chern-Simons vector portal. 
We conclude in Sec.~\ref{sec:conclusion} and provide additional material in Appendices.

\section{Displaced vertices with the muon tracker}
\label{sec:cms-muon-tracker}
\myparagraph{Description of the scheme.}
CMS (compact muon solenoid) is a beam line azimuthally symmetric detector consisting of a solenoid generating the $3.8$ T magnetic field, the inner trackers that allow to reconstruct the momentum of particles produced in the pseudorapidity range $|\eta| < 2.5$ (where $\eta = -\log [\tan(\theta/2)]$ 
and $\theta$ is the polar angle with respect to the anticlockwise-beam direction) and the muon trackers located outside the solenoid~\cite{Chatrchyan:2008aa}.

The muon system is located outside the solenoid and covers the range $|\eta|<2.4$. It is a set of gaseous detectors sandwiched among the layers of the steel flux-return yoke. This allows for a muon to be detected along the track path at multiple points~\cite{Sirunyan:2018fpa}. The magnetic field in the muon system is not uniform, and goes from 2~T in the innermost part down to almost 0~T in the outer part~\cite{CMS:1997iti}. 
\begin{figure}
    \centering
    \includegraphics[width=0.45\textwidth]{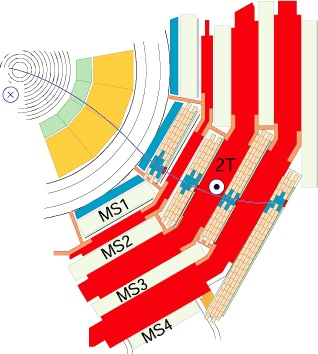}
    \caption{Cross-section of the CMS experiment. Layers (muon stations) of the muon detector in the plane perpendicular to the beam direction. The figure is from~\cite{CMS}.}
    \label{fig:muon-detector}
\end{figure}
Schematic drawing of the muon detector is shown in Fig.~\ref{fig:muon-detector}.

For the LHC Run 2, new reconstruction of muons has been introduced~\cite{CMS_muonreco}, 
the so-called displaced standalone muon reconstruction.
This reconstruction is specifically designed to address cases when muons 
are produced in decays far away from the production vertex. New algorithm achieves 
an almost 100\% reconstruction efficiency for the muon production radius 
up to about 3~m. This is a significant gain in the efficiency compared 
to the reconstruction which uses also inner tracker information, 
but at the same time, the momentum resolution deteriorates 
by about a factor of 10 and is in the range 10--60\%.

The muon tracker can use two muon tracks to reconstruct a displaced vertex 
originating from the decay $X\to \mu\mu+\dots$. The reconstructed DV together with 
the production vertex that can be tagged by prompt decays products, 
\textit{e.g.}\ a prompt lepton, and an underlying event produced together with the $X$ particle,
is identified as a DV event. 
Due to the large distance between a PV and a reconstructed DV, we will call this scheme the ``the long DV'' scheme. 

A presence of a prompt decay product in the event is essential for  event triggering, so that it is recorded and can be analyzed offline.
It should be noticed, that after the phase II upgrade, during the high-luminosity LHC phase, the possibility to use track-trigger in CMS will be introduced. This will enable a possibility 
to reconstruct and identify displaced tracks online~\cite{CMS:2018lqx,CMS:2018qgk,Atlas:2019qfx}, and hence will remove a need
for a prompt lepton in the event. 
However, for the models discussed in this paper, current hardware configuration of the CMS allows to perform the searches with the already recorded data, as well with the data to be obtained during the Run 3 of the LHC.

At the same time, final states with a prompt, well identified, object in the event, 
as \textit{e.g.}\ a prompt muon or electron, have much lower background rate. 
In this case the instrumental backgrounds and non-muon backgrounds from cosmic rays are reduced to a negligible level. The remaining cosmic-ray muon backgrounds 
can be suppressed by selections which do not impact signal efficiency, as described in Ref.~\cite{Aad:2019kiz}. 
The remaining sources of the background for the long DV scheme 
are processes with a presence of a prompt object (as e.g. W boson production)
accompanied by decays of the SM particles into single muons, 
which give rise to combinatorial two-muon events, and two-muon decays 
of the SM particles (for example, $J/\psi, \rho, \omega$ mesons and the $Z$ boson). 
The most significant displacement of such DV appears in case of two muons originating for a heavy-flavor particle decay ($b$ or $c$ hadrons). As we do not carry out an experimental analysis in this paper, we assume that this background is negligible
if one requires the transverse position of the displaced vertex to be as far as $l_{\text{DV}}>2\text{ cm}$ from the beam collision point, since most of the SM particles decay before reaching this displacement~\cite{Sirunyan:2017nbv}. Under this assumption we lose a part of the efficiency for LLPs with shorter lifetimes, but at the same time we provide a more robust estimate of the potential signal sensitivity.
Because of the position of the muon trackers, the muon events can be reconstructed at the distances $l_{\text{DV}} <3\text{ m}$. The muons can be reconstructed with high efficiency and low misidentification probability if each of them has the transverse momentum $p_{T} > 5\text{ GeV}$~\cite{CMS_mulowpt,CMS_mulowptmisID}.

\begin{figure}[h!]
    \centering
    \includegraphics[width=0.45\textwidth]{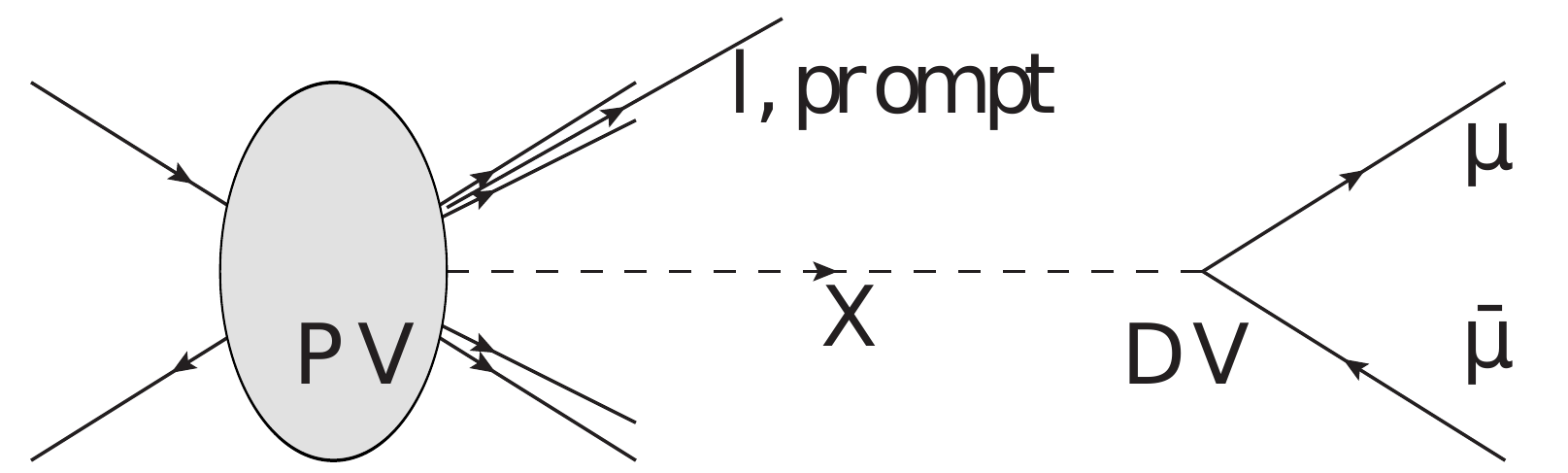}
    \caption{Schematic diagram of the search scheme of LLPs at CMS using the muon detectors. The production vertex (PV) is tagged by the prompt lepton $l$, while the displaced vertex (DV) is reconstructed by two muons produced in the decay $X \to \mu\mu$.}
    \label{fig:displaced-vertex-long-dv}
\end{figure}

To summarize, an event in the long DV search scheme should satisfy the following selection criteria:
\begin{compactitem}[--]
    \item A prompt electron with $|\eta|<2.5, \ p_{T} > 30\text{ GeV}$ or a prompt muon with $|\eta|<2.4, \ p_{T} >25\text{ GeV}$, which are required for an event to be recorded by the single lepton triggers;
    \item The minimal transverse displacement of the DV from the PV is $l_{\text{min},\perp} = 2\text{ cm}$; the maximal transverse and longitudinal displacements are $l_{\text{max},\perp} = 3\text{ m}, \ l_{\text{max},l}= 7\text{ m}$;
    \item Two displaced muon tracks, each with $|\eta|<2.4, \ p_{T} > 5\text{ GeV}$.
\end{compactitem}
The requirement of a large displacement of a DV from the PV helps to significantly reduce the background from SM processes. Therefore, even in the region with the invariant mass of two muons below 5 GeV (mass of $B$-mesons) the SM background is considered to be negligible. The scheme is presented in Fig.~\ref{fig:displaced-vertex-long-dv}.

An event with prompt $\tau$ lepton can be tagged by its leptonic decays $\tau \to l\bar{\nu}_{l}\nu_{\tau}$, where the leptons $l = e/\mu$ satisfy the criteria for prompt leptons presented above. We do not consider the reconstruction of $\tau$ leptons by their hadronic decay products since the trigger threshold for $p_{T}$ of hadronic decay products is too high for efficient reconstruction.\footnote{Current trigger threshold is $p_{T} > 180\text{ GeV}$.} However, in the future it is wise to invest into the development of a dedicated multi-object trigger which would allow to bring down the prompt tau $p_{T}$ by including additional displaced leptons in the event.

In~\cite{Drewes:2019fou} a similar search scheme was discussed, albeit with less restrictive selection criteria  $l_{\text{min},\perp} = 0.5 \text{ cm}$, $l_{\text{max},\perp} = 4\text{ m}$, and $|\eta|<4$ for leptons.\footnote{After the HL-LHC upgrade the CMS will extend its pseudorapidity range to $|\eta| < 4$.} A wider range of muon pseudorapidities leads to the enlargement of the selection efficiency, while a smaller displacement between a DV and the PV lifts up the upper bound of the sensitivity and hence increases the maximal mass reach. However, we caution that the background-free hypothesis for the region with smaller DV displacements adopted in~\cite{Drewes:2019fou} has not been tested.
Nevertheless, to demonstrate potential improvement from considering lower displacements we provide sensitivity for two scenarios: ``realistic'' for the selection criteria outlined above, and ``optimistic'', defined according to~\cite{Drewes:2019fou}.

\myparagraph{Estimation of the number of events.}
The number of decay events of a new particle $X$ that pass the selection criteria is
\begin{equation}
  \label{eq:n-events}
  N_{\events} = N_{\text{parent}} \cdot \Br_{\text{prod}} \cdot P_{\decay}\cdot \epsilon,
\end{equation}
where
\begin{compactitem}[--]
\item $N_{\rm parent}$ is the total number of parent particles that produce a particle $X$ at the LHC;
\item $\Br_{\text{prod}}$ is the branching fraction of the production of a particle $X$ in decays of the parent particle; 
\item $P_{\decay}$ is the decay probability,
\begin{multline}
P_{\decay} = \int d\theta_{X}dp_{
X}f(p_{X},\theta_{X})\times \\ \times \left(e^{-l_{\min}/c\tau_{X}\gamma_{X}}-e^{-l_{\max}/c\tau_{X} \gamma_{X}}\right),
\label{eq:pdecay}
\end{multline}
with $\tau_{X}$ being the proper lifetime of the particle $X$, $ \gamma_X$ is its $\gamma$ factor, and $f(p_{X},\gamma_{X})$ is the distribution function of the $X$ particle whose decay products satisfy the selection criteria; 
\item $\epsilon$ is the \emph{overall efficiency} -- the fraction of all decays of the $X$ particle that occurred in the decay volume between $l_{\min}$ and $l_{\max}$, have passed the selection criteria, and were successfully reconstructed.
\end{compactitem}
The efficiency is a combination of several factors:
\begin{equation}
    \epsilon = \epsilon_{\text{sel}}\cdot \epsilon_{\text{rec}} \cdot \text{Br}_{X\to \mu\mu},
    \label{eq:efficiency}
\end{equation}
where $\epsilon_{\text{sel}}, \epsilon_{\text{rec}}$ are the efficiencies of the selection and subsequent reconstruction of an event correspondingly, and $\text{Br}_{X \to \mu\mu}$ is the branching ratio of the decay of the $X$ particle into two muons. 
Clearly, $\epsilon_{\text{rec}}$ does not depend on the nature of LLP.
The reconstruction efficiency for leptons is well above 95\% for muons with $p_{T} > 5$ GeV~\cite{CMS_mulowpt,CMS_muonreco,Sirunyan:2018fpa} and for electrons with $p_{T} > 30$ GeV~\cite{Khachatryan:2015hwa}. 
Therefore, for simplicity the reconstruction efficiency is taken to be equal to 1 ($\epsilon_{\text{rec}} = 1$) in what follows.

We define the sensitivity curves  by the condition $N_\events \simeq 3$, corresponding to the 95\% exclusion limit under the assumption of zero background. The lower boundary can be easily rescaled to other $N_\events$.

\begin{figure}
    \centering
    \includegraphics[width=0.5\textwidth]{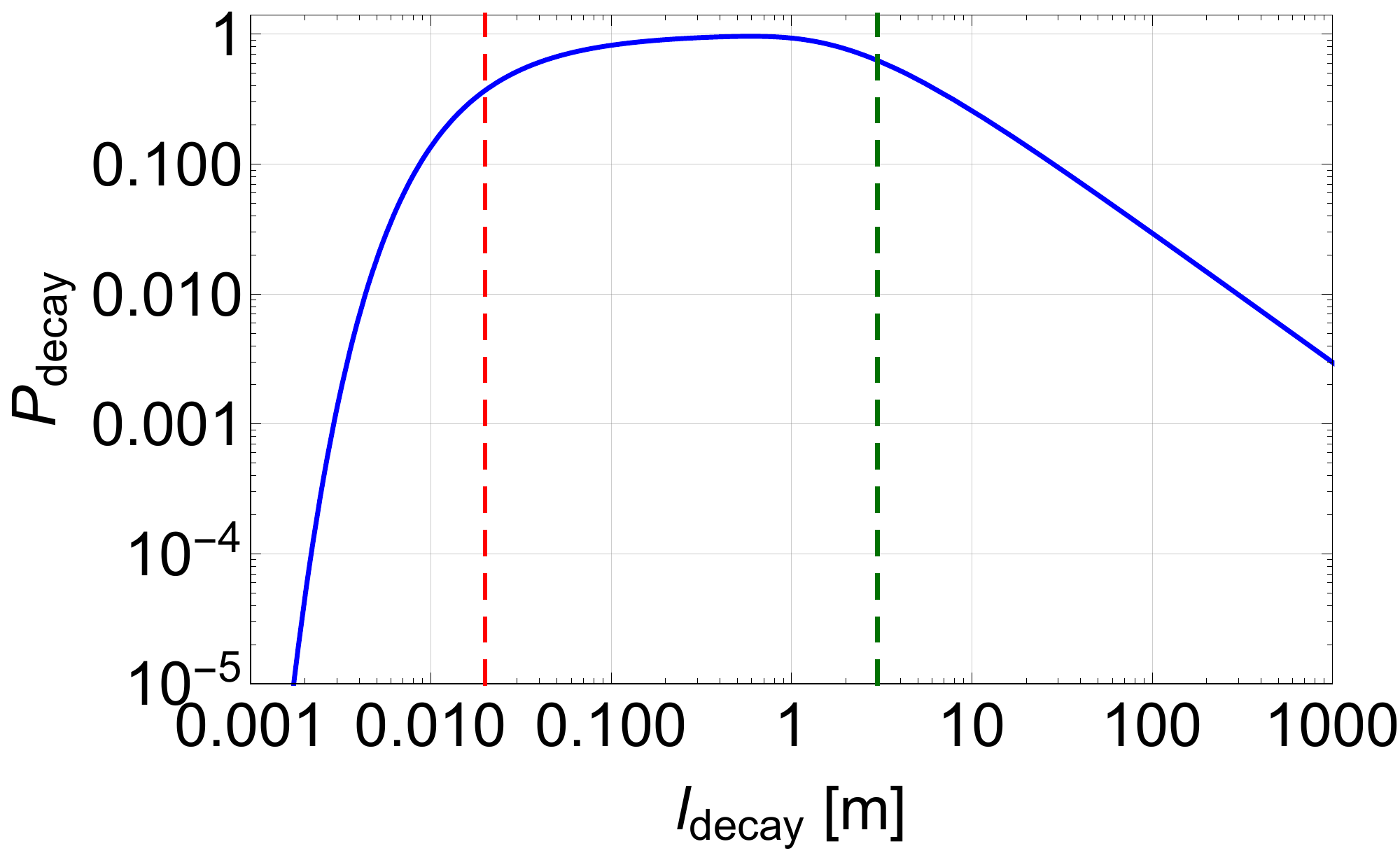}
    \caption{Dependence of the decay probability~\eqref{eq:pdecay} on the decay length $l_{\decay}$. For simplicity we assumed that all the particles travel with the same momentum and in the same direction, and set $l_{\max} = 3\text{ m}$. The dashed lines denote the values $l_{\decay} = l_{\min}$ and $l_{\decay} = l_{\max}$. In the domain $l_{\decay}\gg l_{\max}$ the decay probability scales as $P_{\decay}\simeq l_{\max}/l_{\decay}$, while in the domain $l_{\min}\lesssim l_{\decay}\lesssim l_{\max}$ it behaves approximately constantly and does not depend on $l_{\max}$.}
    \label{fig:decay-probability}
\end{figure}

\myparagraph{Potential of the scheme.}
The main advantage of the long DV scheme is the large length of the fiducial decay volume $l_{\max}$, which exceeds the lengths of the decay volumes of other DV search schemes at the LHC (see, e.g.,~\cite{Cottin:2018nms,Aaij:2014aba}) by $\simeq 10$ times. 
This has a benefit when searching for new particle with large decay lengths,
\begin{equation}
l_{\decay} \equiv c\tau_{X}\gamma_{X} \gg l_{\text{max}}
\label{eq:condition1}
\end{equation}
Indeed, in this case the decay probability~\eqref{eq:pdecay} is in the ``linear regime'', $P_{\decay} \approx l_{\max}/l_{\decay}$, and as a result the number of events~\eqref{eq:n-events} is proportional to $l_{\max}$. For decay lengths that do not satisfy the condition~\eqref{eq:condition1} the decay probability does not depend on $l_{\max}$, and the improvement is lost (see Fig.~\ref{fig:decay-probability}).

\begin{figure}[h!]
    \centering
    \includegraphics[width=0.45\textwidth]{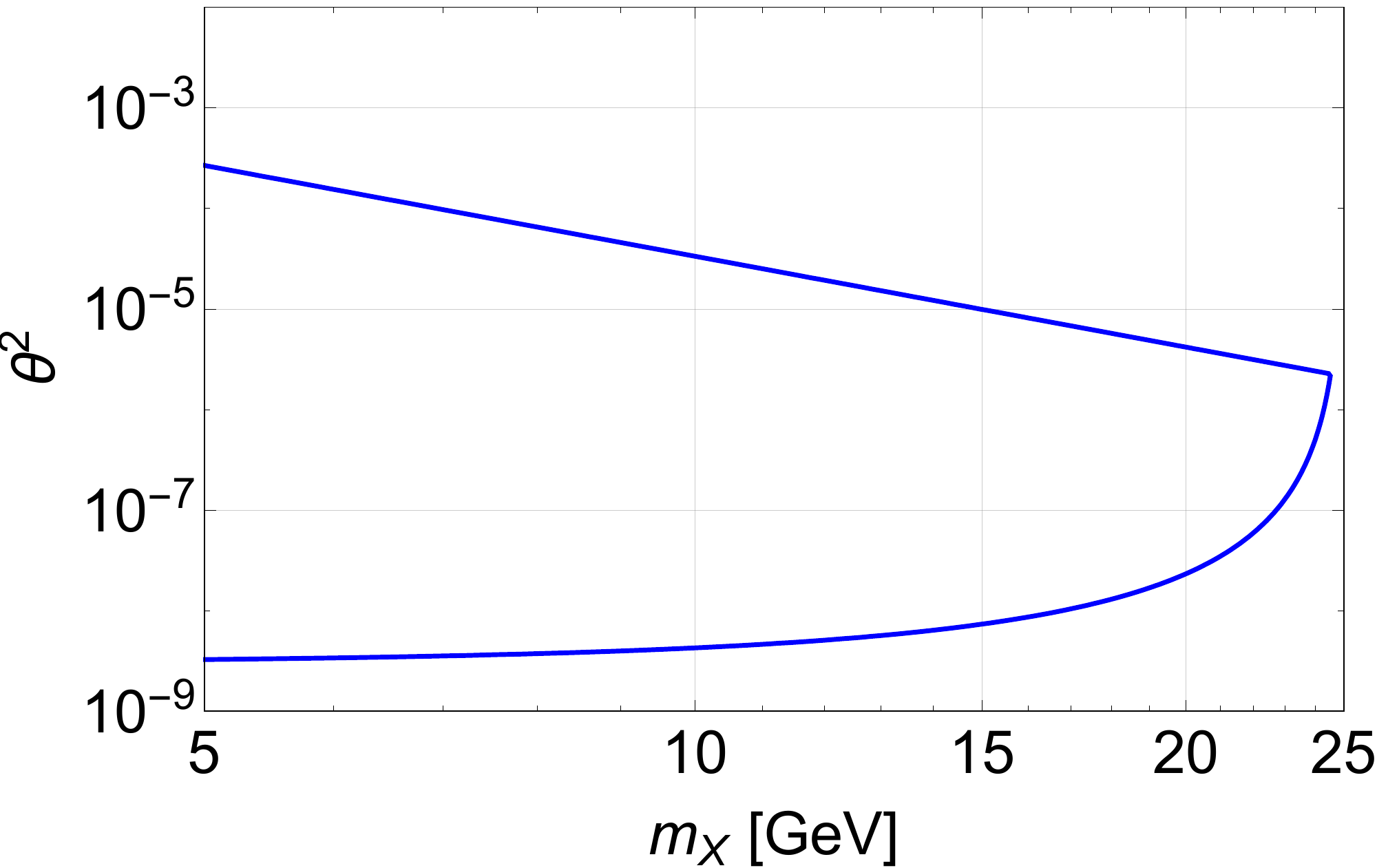}
    \caption{The illustration of the parameter space which is optimal for being probed with the long DV scheme, see text for details. We used a toy model with $N_{\text{prod}} = 10^{9}\left[1-m_{X}^{2}/(25\text{ GeV})^{2}\right]^{2}\theta^{2}_{X}$, $\text{Br}_{X\to \mu\mu} = 1$ and $l_{\decay} = 0.1m_{X}^{-3}\theta_{X}^{-2}\text{ m}$.}
    \label{fig:toy-example}
\end{figure}

In order to probe the domain~\eqref{eq:condition1} there must be sufficient production of the $X$ particles, i.e. 
\begin{equation}
N_{\text{prod}}\cdot\text{Br}_{X\to \mu\mu}>3,
\label{eq:condition2}
\end{equation}
where $N_{\text{prod}} = N_{\text{parent}}\cdot\text{Br}_{\text{prod}}$. The parameter space defined by the conditions~\eqref{eq:condition1},~\eqref{eq:condition2} is optimal for being probed by the long DV scheme. A toy example of the parameter space is given in Fig.~\ref{fig:toy-example}.

\section{Examples of probed models}
\label{sec:examples}
\subsection{HNLs}
Consider the model of heavy neutral lepton (HNL), which introduces a Majorana fermion singlet $N$ coupled to the gauge invariant operator $\bar{L}\tilde{H}$, where $L$ is the left lepton doublet and $\tilde{H} = i\sigma_{2}H^{*}$ is the Higgs doublet in conjugated representation. This coupling introduces the mass mixing between the HNL and active SM neutrino $\nu_{l}$ parametrized by the mixing angle $U_{l}$. We adopt the phenomenology of the HNLs from~\cite{Bondarenko:2018ptm}. The main production channel of the HNLs with masses in the range $m_{N}\gtrsim 5\text{ GeV}$ is the decay of the $W$ bosons. We use the value $\sigma_{W} \approx 190\text{ nb}$ for the total production cross-section of the $W$ bosons at the LHC at energies $\sqrt{s} = 13\text{ TeV}$~\cite{Aad:2016naf}. To estimate the parameter space defined by~\eqref{eq:condition1} and~\eqref{eq:condition2}, we calculated the energy spectrum and geometric acceptance $\epsilon_{\text{geom}}$ of the HNLs in the pseudorapidity range $|\eta|<2.5$ in LO using the model \textit{HeavyN}~\cite{Degrande:2016aje}. We found $\epsilon_{\text{geom}} \approx 0.5$ for the mass range $m_{N}\lesssim 20\text{ GeV}$ and $E_{N} \approx 80\text{ GeV}$. 

\begin{figure}[h!]
    \centering
    \includegraphics[width=0.45\textwidth]{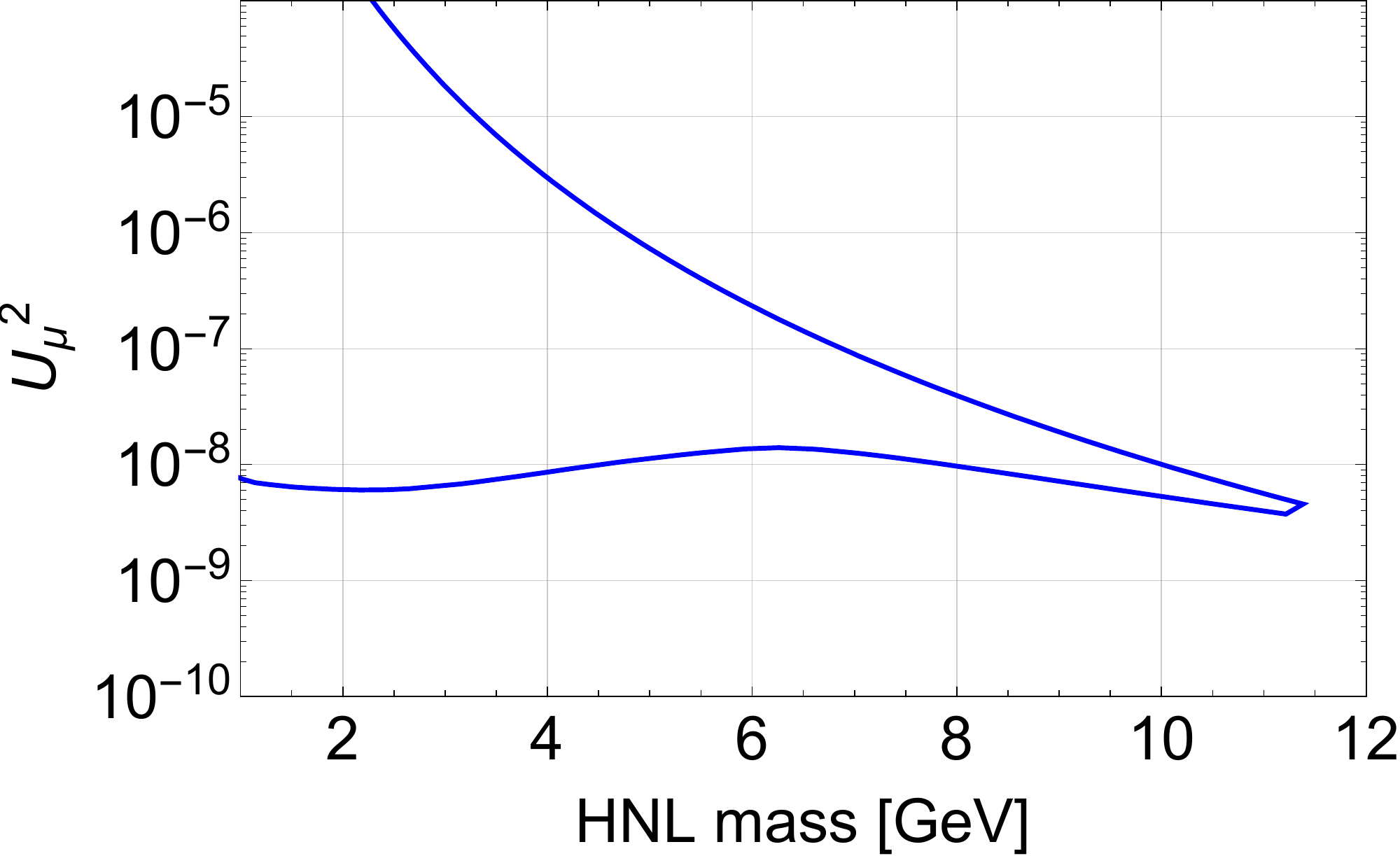}
    \caption{The parameters of HNLs mixing with $\nu_{\mu}$ that satisfy criteria~\protect\eqref{eq:condition1} --\protect\eqref{eq:condition2} for the  LHC luminosity  $\mathcal{L} = 3000\text{ fb}^{-1}$.
    Note, this is \textit{not} an exclusion region, see text around equations for details.
    }
    \label{fig:hnl-parameter-space}
    \end{figure}

In Fig.~\ref{fig:hnl-parameter-space} we show the parameter space for the HNLs mixing with $\nu_{\mu}$ that can be optimally probed by the long DV scheme (see Sec.~\ref{sec:cms-muon-tracker}).
We see that the domain where the long DV scheme has good potential corresponds to the masses $m_{N}<10\text{ GeV}$ and the mixing angles $U^{2}\gtrsim 10^{-9}$. 

\myparagraph{Simulations.}
To find the efficiency for the HNLs mixing with $\nu_{e/\mu}$, we used MadGraph5~\cite{Alwall:2014hca} with the model \textit{HeavyN}~\cite{Degrande:2016aje}. For simulating decays of $\tau$ lepton, we used \textit{taudecay\_UFO} model~\cite{Hagiwara:2012vz}. For the mixing with $\nu_{e/\mu}$ we simulated the process $p + p\to W, \ W \to l + N$, where $l = e$ for the mixing with $\nu_{e}$ and $l = \mu$ for the mixing with $\nu_{\mu}$, with subsequent decay $N \to \mu^{+}+\mu^{-}+\nu/\bar{\nu}_{l}$. In the case of the mixing with $\nu_{\tau}$, we simulated the process $p + p \to W, W \to \tau + N$ with subsequent decays $N \to \mu^{+}+\mu^{-}+\nu_{\tau}/\bar{\nu}_{\tau}$ and $\tau \to l+\bar{\nu}_{l}+\nu_{\tau}$, where $l = e/\mu$. 

Using the selection criteria for the long DV scheme, we computed the selection efficiencies. They were found to be almost independent of the mass of the HNL in the mass range $1\text{ GeV}<m_{N} <20\text{ GeV}$. We give their values in Table~\ref{tab:efficiency-hnl}.  
\begin{table}[]
    \centering
    \begin{tabular}{|c|c|c|c|c|}
        \hline $\epsilon_{\text{sel}}$ & $e$& $\mu$ & $\tau$ \\ \hline
      Realistic & 0.16 & 0.17 & $7\cdot 10^{-3}$ \\ \hline
      Optimistic  & 0.26 & 0.31 &$3.2\cdot 10^{-2}$\\ \hline
    \end{tabular}
    \caption{The values of the selection efficiencies for HNLs of different flavors $e,\mu,\tau$ in the case of realistic and optimistic selection criteria}
    \label{tab:efficiency-hnl}
\end{table} 
The suppression of the efficiency for  mixing with $\nu_{\tau}$ is due mainly to the reconstruction of the prompt $\tau$ event. Indeed, the amount of the leptons produced in the decay $\tau \to l \bar{\nu}_{l} \nu_{\tau}$ and passing the $p_{T}$ selection criterion for the prompt leptons is $\approx 0.1$.

For the average momentum we found $p_{N} \approx 70\text{ GeV}$ and $p_{N} \approx 180 \text{ GeV}$ for the realistic and optimistic estimates correspondingly.  

\begin{figure}[h!]
    \centering
    \includegraphics[width=0.45\textwidth]{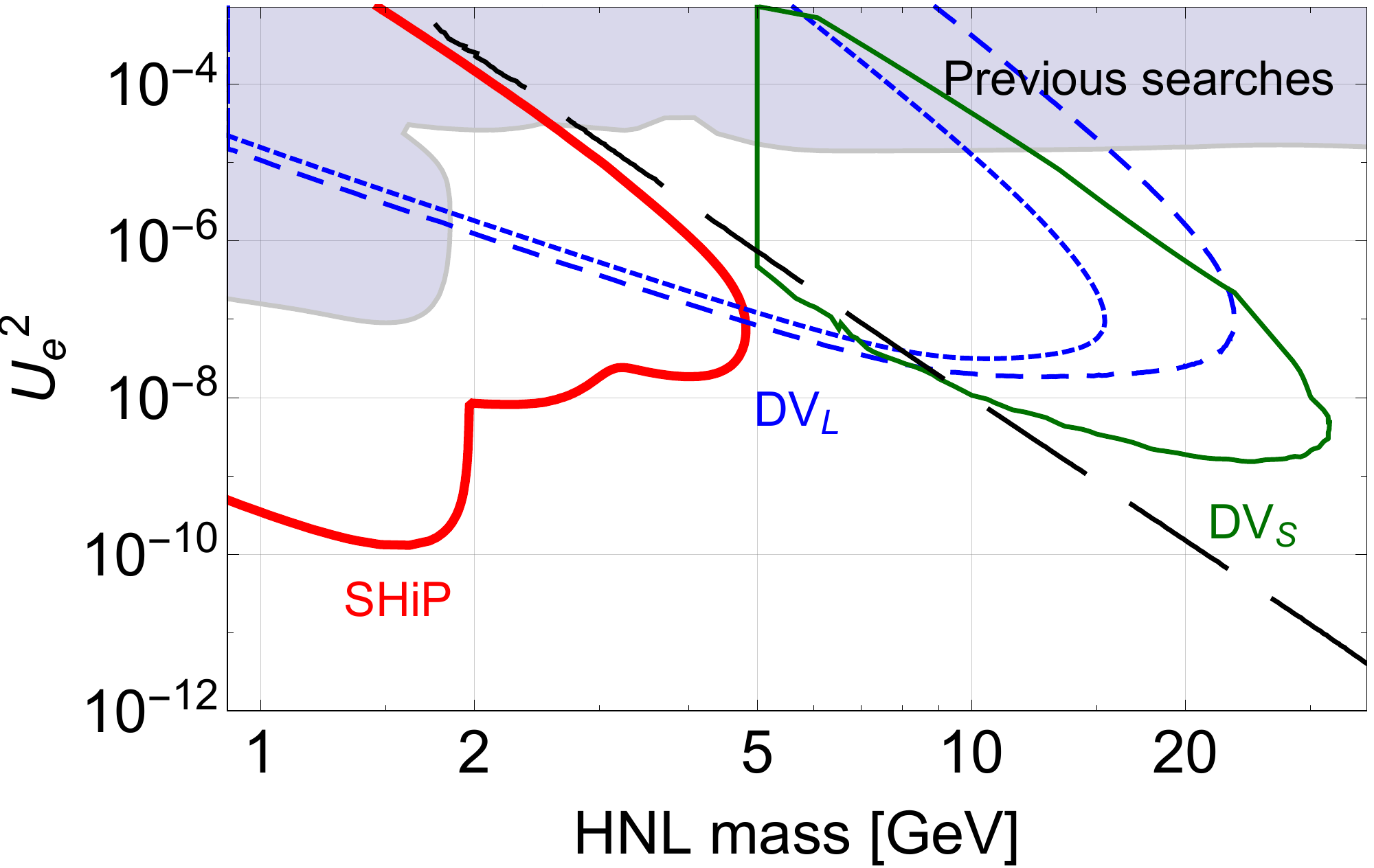}
    \\
    \includegraphics[width=0.45\textwidth]{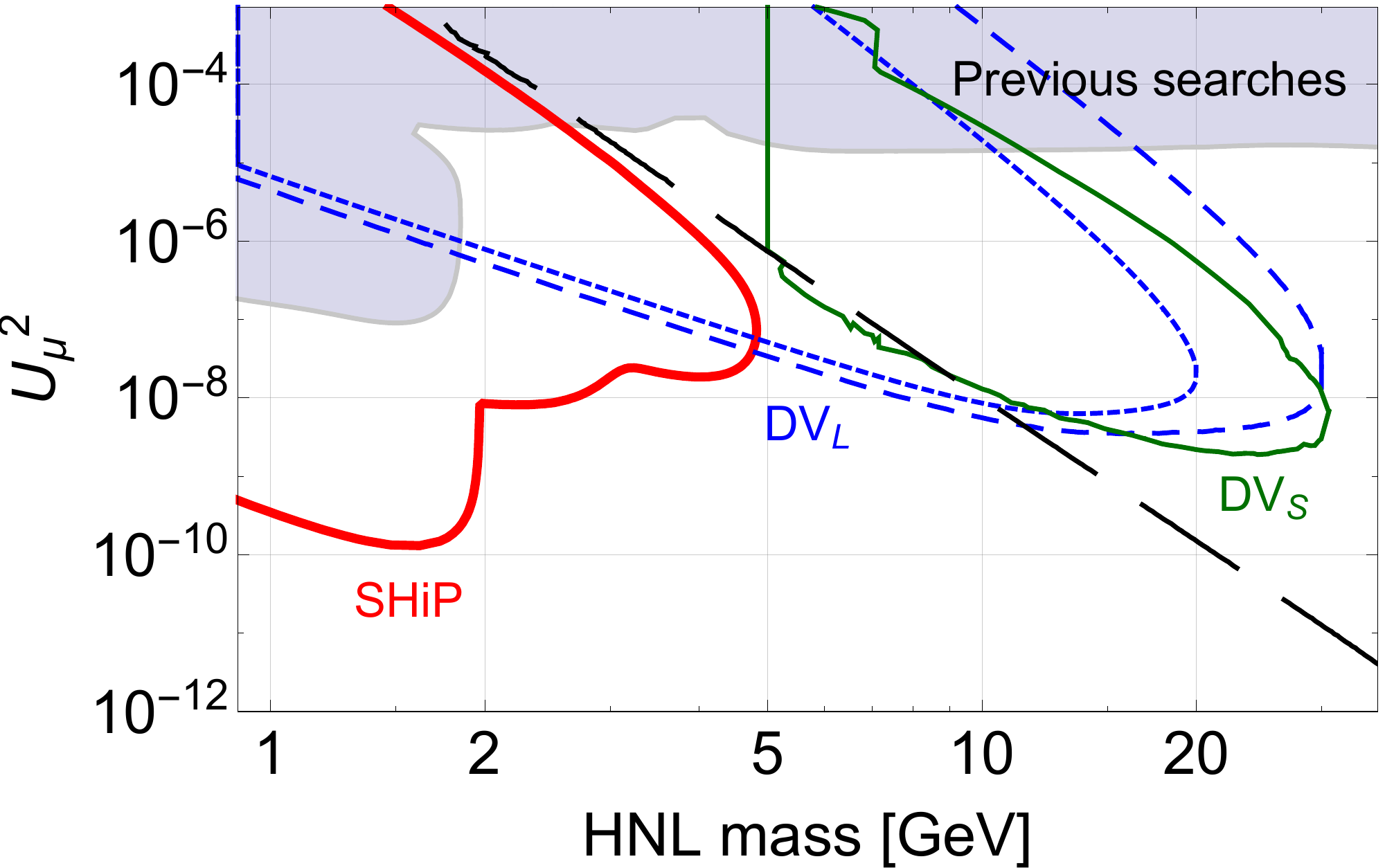}
    \\
    \includegraphics[width=0.45\textwidth]{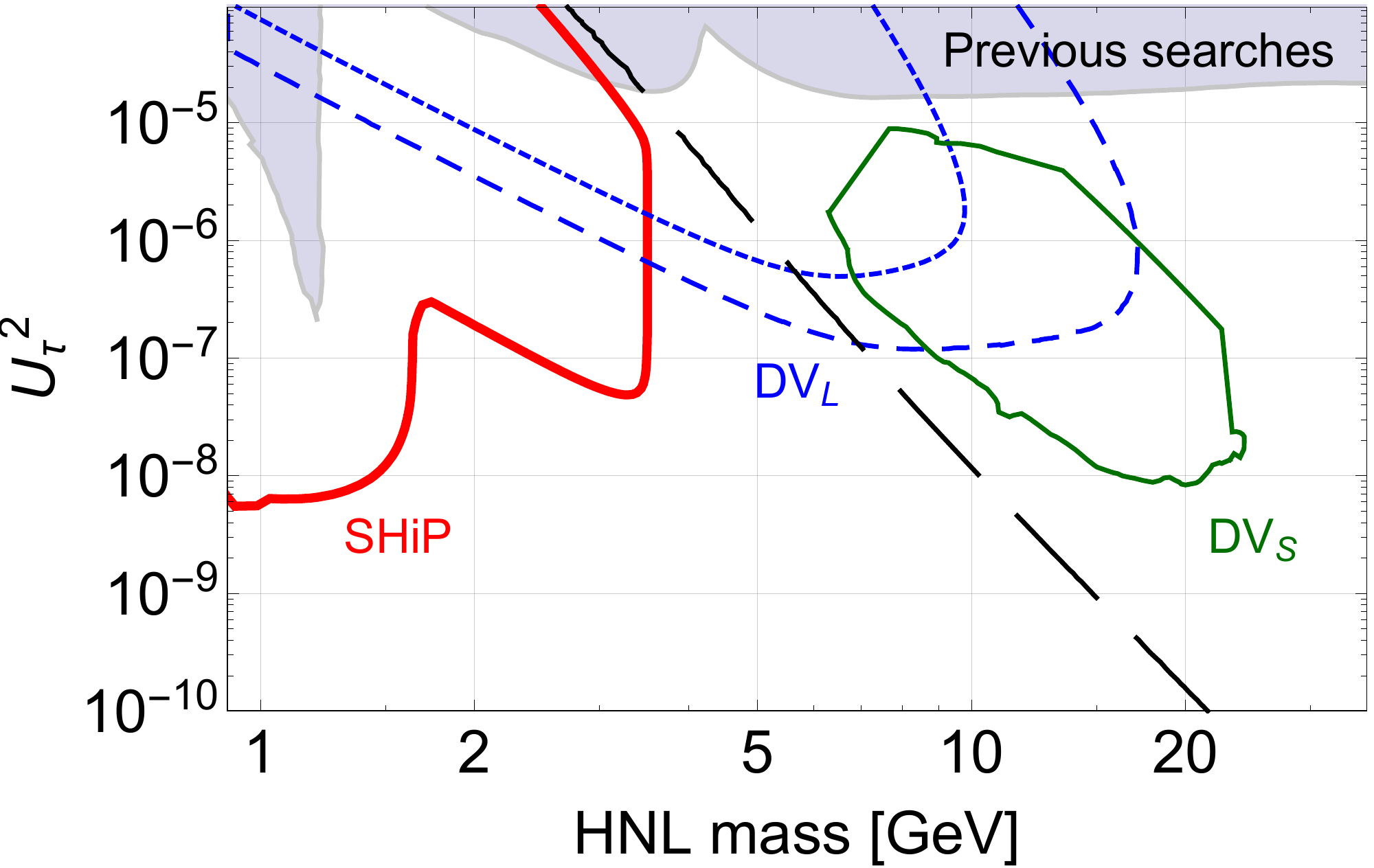}
    \caption{The sensitivity of the long DV ($DV_{L}$) and short DV ($DV_{S}$) search schemes to HNLs mixing with $\nu_{e}$ (upper panel), $\nu_{\mu}$ (middle panel) and $\nu_{\tau}$ (lower panel). By the blue short dashed line we denote the realistic sensitivity obtained using the selection criteria presented in this paper, while the blue dashed line corresponds to the optimistic estimate of the sensitivity using relaxed selection criteria from~\cite{Drewes:2019fou}, see Sec.~\ref{sec:cms-muon-tracker} for details. The sensitivity of the SHiP experiment is taken from~\cite{SHiP:2018xqw}. Black long-dashed line indicates HNL parameters that correspond to $l_{\decay} = 3\text{ m}$. The estimates are for the high luminosity LHC phase, $\mathcal{L} = 3000\text{ fb}^{-1}$. For the DV search schemes sensitivity we require $N_{\text{events}}\ge 3$ and assume zero background (see text for details).} 
    \label{fig:long-short-comparison}
\end{figure}

\myparagraph{Comparison with other schemes.}
Let us compare the sensitivity of the long DV search scheme to the HNLs with a scheme from~\cite{Cottin:2018nms,Cottin:2018kmq} that uses inner trackers at ATLAS to search for DVs events (c.f.~\cite{Boiarska:2019jcw}). 
Owing to its smaller transverse displacement $l_{\max} = 0.3\text{ m}$ we call it the ``\textit{short DV scheme}''. 
For the estimation of the sensitivity of the short DV scheme we use parameterized efficiencies $\epsilon(m_{N},U^{2})$ provided by the authors of~\cite{Cottin:2018nms}. The comparison of the sensitivities is given in Fig.~\ref{fig:long-short-comparison}. We show both optimistic and realistic estimate of the sensitivity of the long DV scheme. We also show the sensitivity of the SHiP experiment from~\cite{SHiP:2018xqw} that serves for an illustration of the sensitivity reach of Intensity Frontier experiments.

The long DV scheme allows to search for HNLs in the unexplored region of the parameter space that is not accessible to other Intensity Frontier experiments or other LHC searches.
Its difference in the sensitivity with the short DV scheme is due to three reasons. First, for masses $m_{N} \lesssim 10\text{ GeV}$ the decay probability for both the schemes is in the linear regime (see Sec.~\ref{sec:cms-muon-tracker}), and therefore the long DV scheme gets the benefit from the $10$ times larger length of the decay volume $l_{\max}$. Second, for the masses $5\text{ GeV}\lesssim m_{N} \lesssim 10\text{ GeV}$ there is a drop of the overall efficiency for the short DV scheme. This is caused by the selection criteria on the reconstructed invariant mass of the DV, $m_{\text{DV}} > 5\text{ GeV}$, and the charged tracks, $N_{\text{trk}} \geqslant 4$, that are needed to remain in the background free region~\cite{Aaboud:2017iio}. Third, because of absence of the hadronic background the long DV scheme can probe the parameter space $m_{N} \lesssim 5\text{ GeV}$, which is not reachable by the short DV scheme. 

Nevertheless, both the schemes are complementary to each other and provide a cross-check in the mass region $5\text{ GeV}< m_{N}<15\text{ GeV}$. 

\subsection{Chern-Simons portal}
Chern-Simons portal introduces a vector particle $X$ interacting with pseudo-Chern-Simons current of the SM gauge bosons~\cite{Antoniadis:2009ze,Alekhin:2015byh}:
\begin{multline}
    \mathcal{L}_{CS} = c_W\epsilon^{\mu\nu\lambda\rho} X_\mu W_\nu \partial_\lambda W_\rho +\\
  c_{\gamma} \cos\theta_W \epsilon^{\mu\nu\lambda\rho} X_\mu Z_\nu \partial_\lambda \gamma_\rho +
  c_{Z} \sin\theta_W \epsilon^{\mu\nu\lambda\rho} X_\mu Z_\nu \partial_\lambda Z_\rho
  \label{eq:chern-simons}
\end{multline}
We can add the interaction of the $X$ boson with SM leptons in the form
\begin{equation}
    \mathcal{L}_{X\mu\mu} = c_{W}g_{Xll} X^{\nu}\sum_{l = e,\mu,\tau}\bar{l}\gamma_{5}\gamma_{\nu}l,
    \label{eq:interaction-with-muons}
\end{equation}
where $g_{Xll}$ is a dimensionless constant.\footnote{The coupling $g_{Xll}$ can be generated effectively by the interaction~\eqref{eq:chern-simons} or be an effect of new physics.} 

Let us consider the case when $c_{\gamma},c_{Z}\ll c_{W}$. Then the production of the $X$ particle in $pp$ collisions goes through the $XWW$ vertex, while the decay goes through the vertex~\eqref{eq:interaction-with-muons} down to very small couplings $g_{Xll}^{2} \simeq 10^{-7}$ for the $X$ bosons as heavy as $m_{X} \simeq 40\text{ GeV}$, see Appendix~\ref{sec:CS}. These vertices are parametrically independent, and for particular values of $g_{Xll}$ it is possible to probe the parameter space in the optimal domain for the long DV scheme. The process of interest is
\begin{equation}
W \to X +l+ \bar{\nu}_{l}, \quad X \to \mu^{+}+\mu^{-}
\end{equation}
The lepton $l$ produced in the $W$ decay can be triggered as a prompt lepton, while the muon pair from the decay of the $X$ boson can be reconstructed as displaced muons, which meets the selection criteria of a DV event within the long DV scheme.

To find the selection efficiency and the energy spectrum of the $W$ bosons, we implemented the model of the $X$ boson~\eqref{eq:chern-simons},~\eqref{eq:interaction-with-muons} into the MadGraph using FeynRules~\cite{Alloul:2013bka,Christensen:2008py}. 
The model is publicly available~\cite{ChernSimonsFeynRules}. 
We simulated the processes $p + p \to e^{+}/\mu^{+}+\nu_{e/\mu} +X$ (plus the charge conjugated final states) with subsequent decays $X \to \mu^{+}\mu^{-}$. We have found that the overall efficiency is $\epsilon \approx 2.3\cdot 10^{-2}$ for $m_{X}$ ranging from $1\text{ GeV}$ to $20\text{ GeV}$. The average momentum of the $X$ boson $p_{X} \approx 40\text{ GeV}$.

\begin{figure}[h!]
    \centering
    \includegraphics[width=0.45\textwidth]{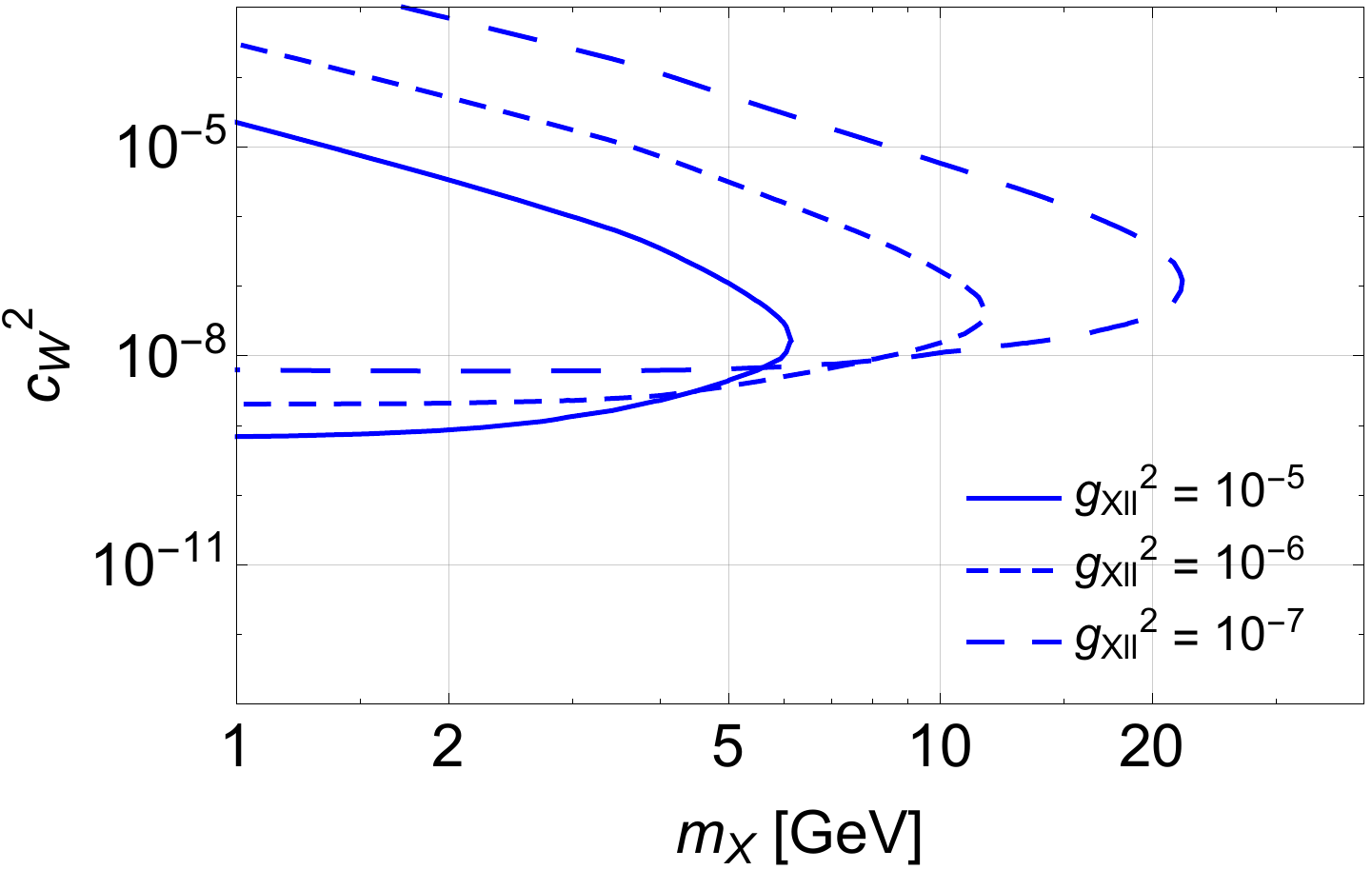}
    \caption{The sensitivity of the long DV scheme at the high luminosity phase to the Chern-Simons portal for different values of the coupling to muons~\eqref{eq:interaction-with-muons}. For the DV search schemes sensitivity we require $N_{\text{events}}\ge 3$ and assume zero background (see text for details).}
    \label{fig:CS}
\end{figure}

The sensitivity to the Chern-Simons portal is shown in Fig.~\ref{fig:CS}. We conclude that the long DV scheme can probe masses up to $m_{X} \simeq 30\text{ GeV}$ and couplings down to $c_{W}^{2}\simeq 10^{-9}$. We note that the probed parameter space is well below the current experimental bound on $c_{W}$, which is $c_{W}^{2}\lesssim 10^{-3}(m_{X}/1\text{ GeV})^{2}$~\cite{Alekhin:2015byh}.

\section{Conclusion}
\label{sec:conclusion}
In this paper we proposed a new method of searching for long-lived particles at LHC (``\textit{the long DV scheme}'') that utilizes  the muon tracker at the CMS experiment. 
It uses a prompt lepton and a displaced muon pair to reconstruct a displaced vertex event. The scheme is optimal for probing the parameter space of the LLPs with the decay lengths $l_{\decay} \gtrsim 3\text{ m}$. We demonstrated the potential of the scheme using two exemplary models: heavy neutral lepton (HNL) and Chern-Simons portal. For the HNLs we made a comparison between the long DV scheme and other planned searching schemes at ATLAS/CMS, see \textit{e.g.}~\cite{Cottin:2018nms,Boiarska:2019jcw}.

Our conclusions are the following:
\begin{itemize}[--]
\item For the HNLs, the long DV scheme can probe the parameter space in the mass range $m_{N} \lesssim 20\text{ GeV}$ and down to the mixing angles $U^{2} \sim 10^{-8}$ (when mixing with $\nu_\mu$).
\item The long DV scheme has a unique opportunity to probe the LLPs that decay predominantly into leptons, which is demonstrated by the example of the Chern-Simons portal; 
\item The long DV serach scheme has a sufficiently low SM background even for LLPs with the masses $m \lesssim 5\text{ GeV}$, which is unavailable for DV search schemes at the LHC that look for hadronic decay products. 
In the case of HNLs this gives a possibility to probe the parts of the parameter space that have not been probed by previous experiments and are outside the reach of  the planned Intensity Frontier experiments.
\end{itemize}

\subsubsection*{Acknowledgements.}
This project has received funding from the European Research Council (ERC) under the European Union's Horizon 2020 research and innovation programme (GA 694896,  758316) and from the Netherlands Science Foundation (NWO/OCW).

\newpage
\appendix
 
\section{Chern-Simons portal}
\label{sec:CS}
\myparagraph{Decay width of the $W$ boson into $X$ boson.}
\begin{figure}[h!]
    \centering
    \includegraphics{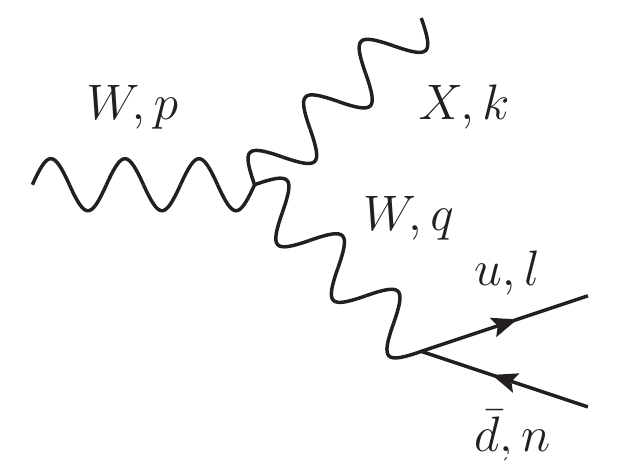}
    \caption{The diagram of the $W$ boson decay into the $X$ boson that goes through the $XWW$ vertex in~\eqref{eq:chern-simons}.}
    \label{fig:w-decay-x}
\end{figure}
The matrix element of the process shown in Fig.~\ref{fig:w-decay-x} is
\begin{multline}
    \mathcal{M}_{W\to X+f+f'}=\frac{g c_{W}}{2\sqrt{2}}\varepsilon^{\mu\nu\alpha\beta}\epsilon_{\mu}(p)(2p+k_{X})_{\nu}\epsilon^{*}_{\alpha}(k_{X})\times \\ \times D_{\beta\gamma}(k_{f}+k_{f'}) \bar{u}(k_{f})\gamma^{\gamma}(1-\gamma_{5})v(k_{f'}),
\end{multline}
where $\epsilon(p)$ is the polarization vector.
In the limit $m_{X,f,f'}\ll m_{W}$ the decay width is
\begin{equation}
    \Gamma_{W\to X+f+f'}\approx c_{W}^{2}\frac{\alpha_{W}m_{W}^{3}}{432\pi^{2}m_{X}^{2}}
\end{equation}
Using this decay width, for the branching ratio of the process $W \to X + l + \nu_{l}$, where $l = e,\mu$ we obtain
\begin{equation}
    \text{Br}_{W\to X + l + \nu_{l}} \approx 2.9 \ (1\text{ GeV}/m_{X})^{2}c_{W}^{2}
\end{equation}
\myparagraph{Tree-level decays generated by $c_{W}$ coupling.} The vertex $c_{W}$ generates the tree level decay process $X \to ff' f'' f'''$. Based on dimensional grounds, we can estimate the corresponding decay width as
\begin{equation}
\Gamma_{X\to ff'f''f'''} \sim c_{W}^{2}\frac{\alpha_{W}^{2}}{(2\pi)^{3}}\frac{m_{X}^{9}}{m_{W}^{8}}
\end{equation}
For the ratio of this decay width to the decay width into two muons, $\Gamma_{X\to \mu\mu} \sim c_{W}^{2}g_{X\mu\mu}^{2}m_{X}/(2\pi)$, we have 
\begin{equation}
\frac{\Gamma_{X \to ff' f'' f'''}}{\Gamma_{X \to \mu\mu}} \simeq \left(\frac{\alpha_{W}}{2\pi g_{X\mu\mu}}\right)^{2}\left( \frac{m_{X}}{m_{W}}\right)^{8}\simeq \frac{10^{-7}}{g_{X\mu\mu}^{2}}\left(\frac{m_{X}}{40\text{ GeV}}\right)^{8},
\end{equation}
which means that for $m_{X} \lesssim 40\text{ GeV}$ the process $X\to ff'f''f'''$ is not relevant for $g_{X\mu\mu}^{2}>10^{-7}$.
\bibliographystyle{JHEP}
\bibliography{ship}

\end{document}